\documentclass[a4paper,11pt]{article}

\usepackage[top=1in,bottom=1in,left=1in,right=1in]{geometry}
\usepackage{amsmath,amssymb,amsthm}
\usepackage{microtype}
\usepackage{cite}
\usepackage{hyperref}
\hypersetup{hidelinks}
\numberwithin{equation}{section}
\allowdisplaybreaks[2]
\theoremstyle{plain}
\newtheorem{theorem}{Theorem}[section]

\newtheorem{lemma}[theorem]{Lemma}

\theoremstyle{definition}

\newtheorem{example}[theorem]{Example}

\theoremstyle{remark}

\newcommand{\GL}{\operatorname{GL}}
\newcommand{\Gr}{\operatorname{Gr}}
\newcommand{\SL}{\operatorname{SL}}
\newcommand{\SU}{\operatorname{SU}}

\newcommand{\tr}{\operatorname{tr}}
\newcommand{\diag}{\operatorname{diag}}
\newcommand{\rank}{\operatorname{rank}}

\title{%
	\fontsize{14.4}{20}\selectfont\bfseries 
	Darboux transformation for the ASDYM equations: \\ Solitons and instantons
}
\author{Shangshuai Li$^{1,2}$\thanks{Email: lishangshuai@nbu.edu.cn}
	\\[5pt]
	{\small $^1$ School of Mathematics and Statistics, Ningbo University, Ningbo 315211, China} \\
	{\small $^2$ Graduate School of Mathematics, Nagoya University, Nagoya 464-8602, Japan} 
}
\date{\today}

\begin{document}
\maketitle

\begin{abstract}
	Using a one-fold Darboux transformation for the
	anti-self-dual Yang--Mills equations,
	we express the Yang--Mills Lagrangian density
	in terms of $\tau$-functions.
	For solitons constructed from exponential functions, we show
	that the associated $\tau$-functions admit a
	parametrization by arbitrary points in the Grassmannian $\Gr(N,M;\mathbb C)$,
	and construct the corresponding
	Yang--Mills fields.
	For instantons arising from the
	Atiyah--Drinfeld--Hitchin--Manin (ADHM)
	construction, we reinterpret the ADHM construction in terms of a one-fold Darboux transformation, recover Osborn's formula for the Yang--Mills Lagrangian density, and obtain the corresponding gauge connection.
	
\end{abstract}



\section{Introduction}

The Kadomtsev--Petviashvili II (KP-II) equation is a
nonlinear dispersive wave equation in two spatial dimensions \cite{KadomtsevPetviashvili1970},
given by
\begin{equation}
	(-4u_t+6uu_x+u_{xxx})_x+3u_{yy}=0.
	\label{eq:kp-ii-equation}
\end{equation}
It is well known that a class of soliton solutions of the KP-II equation can be constructed from
points in the real Grassmannian
$\Gr(N,M;\mathbb R)$ \cite{Kodama2017}.
Let $A=(a_{ij})$ be an $N\times M$ full-rank
matrix representing such a point, and $\kappa_1<\cdots<\kappa_M\in\mathbb{R}$ be ordered
spectral parameters. Define
\begin{equation}
	f_i(x,y,t)
	=\sum_{j=1}^M a_{ij}e^{\theta_j},
	\qquad
	\theta_j:=\kappa_jx+\kappa_j^2y+\kappa_j^3t.
	\label{eq:kp-seeds-and-phases}
\end{equation}
The corresponding $\tau$-function admits the
Wronskian representation \cite{FreemanNimmo1983}
\begin{equation}
	\tau(x,y,t)
	=\operatorname{Wr}(f_1,\ldots,f_N)
	=\det(AEV^T),
	\label{eq:kp-wronskian-tau}
\end{equation}
where $E=\diag(e^{\theta_1},\ldots,e^{\theta_M})$ is a diagonal matrix and $V=(\kappa_j^{i-1})_{1\leq i\leq N,\,
	1\leq j\leq M}$
is an $N\times M$ Vandermonde matrix.
Then $u=2\partial_x^2\ln\tau$ solves the KP-II equation.
Applying the Cauchy--Binet formula to
\eqref{eq:kp-wronskian-tau} yields
\begin{equation}
	\tau(x,y,t)
	=\sum_{I\in\binom{[M]}{N}}
	\Delta_I(A)E_IK_I,
	\qquad
	E_I:=e^{\,\sum_{j\in I}\theta_j},
	\qquad
	K_I:=\prod_{1\leq p<q\leq N}
	(\kappa_{i_q}-\kappa_{i_p}),
	\label{eq:kp-cauchy-binet}
\end{equation}
where $[M]=\{1,\ldots,M\}$ is an index set,
$I=\{i_1<\cdots<i_N\}$ selects $N$ components from $[M]$,
and $\Delta_I(A)$ denotes the Pl\"ucker coordinate
indexed by $I$.
This expansion gives a Grassmannian parametrization
of the soliton solutions of the KP-II equation \cite{ChakravartyKodama2009}.
In particular, if $A$ represents a point in the
totally nonnegative Grassmannian
$\Gr_{\geq0}(N,M;\mathbb R)$, so that
$\Delta_I(A)\geq0$ for all $I$, then
$\tau(x,y,t)>0$ globally on $\mathbb R^3$ \cite{KodamaWilliams2013}.
Thus total nonnegativity ensures the global
regularity of these solutions, while the
combinatorial structure of the Grassmannian
describes the
asymptotic line-soliton patterns and their
interactions \cite{ChakravartyKodama2008,KodamaWilliams2014}.

Related connections between $\tau$-functions,
Grassmannian geometry, and soliton interactions
also occur in the two-dimensional Toda lattice
and the defocusing Davey--Stewartson II equation
\cite{BiondiniWang2010,BiondiniKireyevMaruno2022,MarunoBiondini2004,
	NavarroBetancourtEtAl2026}.
A natural question is whether a similar structure
can be identified in higher-dimensional integrable
systems.

The anti-self-dual Yang--Mills (ASDYM) equations
in four dimensions
provide a natural setting for
this question \cite{Yang1977}. 
The ASDYM equations play a fundamental role in
gauge theory and integrable systems, and many
lower-dimensional integrable equations arise from
them through symmetry reductions
\cite{Ward1985,MasonWoodhouse1996}.
Numerous solution-generating methods have been developed
for the ASDYM equations, including 
Riemann--Hilbert problems, B\"acklund transformations,
inverse-scattering techniques, Hirota bilinear methods and Darboux transformations (DTs)
\cite{MasonWoodhouse1996,CorriganFairlieYatesGoddard1978,
	NimmoGilsonOhta2000,UenoNakamura1982,SasaOhtaMatsukidaira1998,BelavinZakharov1978,OhtaNimmoGilson2001}.
In particular,
Darboux transformations yield quasi-Wronskian and
quasi-Grammian solutions expressed in terms of
quasideterminants
\cite{GilsonHamanakaHuangNimmo2020,LiHamanakaHuangZhang2025}. Asymptotic and exact equivalence between the two solution structures was established for the $\GL(2,\mathbb{C})$ ASDYM equations \cite{Li2026,Huang2026}.

Throughout, we use complex coordinates
$(z,w,\widetilde z,\widetilde w)$ with metric
$\mathrm ds^2=2(\mathrm dz\,\mathrm d\widetilde z
-\mathrm dw\,\mathrm d\widetilde w)$,
and adopt the conventions
$\mathcal F=\mathrm d\mathcal A+\mathcal A\wedge\mathcal A$
and
$\mathcal L=-\frac14\tr(\mathcal F_{\mu\nu}\mathcal F^{\mu\nu})$
for the curvature and Yang--Mills Lagrangian density.
In Yang's gauge, the ASDYM equations reduce to
the Yang equation \cite{Yang1977},
\begin{equation}
	\partial_{\widetilde z}
	\bigl((\partial_z\mathcal J)\mathcal J^{-1}\bigr)
	-
	\partial_{\widetilde w}
	\bigl((\partial_w\mathcal J)\mathcal J^{-1}\bigr)
	=0,
	\label{eq:yang-equation}
\end{equation}
where $\mathcal J$ is a $\GL(r,\mathbb C)$-valued
function called the Yang matrix.
The corresponding gauge connection and Yang--Mills density are
\begin{align}
	\mathcal A
	&=-(\partial_z\mathcal J)\mathcal J^{-1}\,\mathrm dz
	-(\partial_w\mathcal J)\mathcal J^{-1}\,\mathrm dw,
	\label{eq:yang-gauge-potentials}\\
	\mathcal L[\mathcal J]
	&=\tr\Bigl[
	\bigl(
	\partial_{\widetilde z}
	\bigl((\partial_z\mathcal J)\mathcal J^{-1}\bigr)
	\bigr)^2
	-
	\partial_{\widetilde w}
	\bigl((\partial_z\mathcal J)\mathcal J^{-1}\bigr)
	\,
	\partial_{\widetilde z}
	\bigl((\partial_w\mathcal J)\mathcal J^{-1}\bigr)
	\Bigr].
	\label{eq:ym-density-yang-matrix}
\end{align}

Using the quasi-Wronskian representation for the
Yang matrix, recent studies have revealed
the asymptotic behavior and multi-soliton scattering
of \eqref{eq:ym-density-yang-matrix}
\cite{HamanakaHuang2020,HamanakaHuang2022}.
More recently, this Yang–Mills density has been expressed in terms of \(\tau\)-functions for a class of ASDYM solutions  \cite{McLoughlin2026},
\begin{equation}
	\mathcal L
	=-\bigl(
	\partial_z\partial_{\widetilde z}
	-\partial_w\partial_{\widetilde w}
	\bigr)^2\ln\tau(z,\widetilde{z},w,\widetilde{w}).
	\label{eq:asdym-tau-density}
\end{equation}
It is therefore natural to ask whether these $\tau$-functions admit the determinant form \eqref{eq:kp-wronskian-tau} for arbitrary Grassmannian points, and whether the corresponding Yang matrix can be constructed explicitly from the same data.\footnote{Since \eqref{eq:asdym-tau-density} is defined only
	where $\tau\neq 0$, throughout the remainder of this paper
	we restrict our attention to domains on which
	$\tau$ is nonvanishing.}

Besides the soliton solutions constructed from exponential
functions, the ASDYM equations are also
well known for their instanton solutions on Euclidean space. The first example is the
Belavin--Polyakov--Schwartz--Tyupkin (BPST) instanton
\cite{BelavinPolyakovSchwartzTyupkin1975},
which was subsequently generalized to multi-instanton
solutions by 't Hooft and by Jackiw, Nohl, and Rebbi (JNR)
\cite{JackiwNohlRebbi1977}.
In particular, the JNR solutions contain the 't Hooft
multi-instantons as a limiting case.
The Atiyah--Ward ansatz provides another construction
based on twistor geometry \cite{AtiyahWard1977},
and their relation to B\"acklund transformations
was studied in \cite{CorriganFairlieYatesGoddard1978}.

A general algebraic construction of instantons
was established by Atiyah, Drinfeld, Hitchin, and Manin
(ADHM) \cite{AtiyahHitchinDrinfeldManin1978}.
It has been proved that the ADHM construction describes all regular finite-action
$\mathrm{SU}(r)$ instantons of arbitrary topological charge,
including the BPST, 't Hooft, and JNR solutions.
Remarkably, the corresponding Yang--Mills density also
admits a determinant representation given by Osborn's formula
\cite{Osborn1981,CorriganGoddardOsbornTempleton1979,DoreyHollowoodKhozeMattis2002},
\begin{equation}
	\mathcal L=-\frac14
	\left(\partial_{x^1}^2+\partial_{x^2}^2+\partial_{x^3}^2+\partial_{x^4}^2\right)^2\ln\det \mathsf R(x),
	\qquad
	x\in\mathbb{R}^4,
	\label{eq:adhm-osborn-density-cartesian}
\end{equation}
where $\mathsf R(x)$ is the matrix determined by the ADHM
factorization condition.
The similarity between
\eqref{eq:asdym-tau-density} and
\eqref{eq:adhm-osborn-density-cartesian}
suggests a connection between the 
$\tau$-functions arising from DTs
and the ADHM construction.
It is of interest to understand how these different constructions are related by a unified framework.

Our main results are as follows.

\begin{enumerate}
	
	\item \textit{Determinant formula for Yang--Mills Lagrangian densities.}
	We derive \eqref{eq:asdym-tau-density} directly from a
	one-fold Darboux transformation and define the associated
	ASDYM $\tau$-function 
	(Theorem~\ref{thm:onefold-darboux-density}).\footnote{A one-fold DT is sufficient for our purposes,
		as an $n$-fold DT for $\GL(r,\mathbb C)$ ASDYM equations
		can be realized as a one-fold DT
		of size $nr$; see Appendix~\ref{app:onefold-nfold-darboux-correspondence} for details.}
	This provides a unified starting point for studying
	soliton solutions and ADHM instantons.

\item \textit{Soliton solutions parametrized by Grassmannian points.}
For arbitrary $[A]\in\Gr(N,M;\mathbb C)$ and
distinct nonzero spectral parameters
$\kappa_1,\ldots,\kappa_M\in\mathbb C$,
we construct an $M\times M$ matrix $\Theta$ such that
\[
\prod_{1\leq i<j\leq M}
(\kappa_j-\kappa_i)\det\Theta
=
\det(AEV^T)
=
\tau,
\]
where $\tau$ has the form
\eqref{eq:kp-wronskian-tau}
(Theorem~\ref{thm:grassmannian-parametrization}).
Moreover, for irreducible $A$, we explicitly construct
a $\SL(r,\mathbb C)$-valued Yang's matrix  with
$r\leq 1+\min\{N,M-N\}$,
which reproduces the same Yang--Mills density
as in Theorem~\ref{thm:grassmannian-parametrization} (Theorem~\ref{thm:grassmannian-recovery}).

	\item \textit{ADHM instantons and Osborn's formula.}
	For ADHM construction describing charge-$k$
	$\SU(r)$ instantons, we construct an
	$(r+3k)\times(r+3k)$ matrix $\Theta$
	satisfying the vacuum linear system
	associated with a one-fold DT.
	It gives $\det\Theta=\det\mathsf R$ and
	thereby recovers Osborn's formula
	\eqref{eq:adhm-osborn-density-cartesian}
	(Theorem~\ref{thm:adhm-darboux-density-equivalence}).
	We further show that the ADHM
	connection can be recovered in this setting through a
	gauge transformation
	(Theorem~\ref{thm:adhm-darboux-connection-recovery}),
	\begin{equation*}
		\mathsf G^{-1}
		{\mathcal A}_{\mathrm{DT}}\mathsf G
		+\mathsf G^{-1}\mathrm d\mathsf G
		=
		\begin{pmatrix}
			0&
			*&
			*\\
			0&
			\mathcal{A}_{\mathrm{ADHM}}&
			*\\
			0&0&0
		\end{pmatrix},
	\end{equation*}
	where ${\mathcal A}_{\mathrm{DT}}$ is the gauge connection associated with the one-fold DT in Theorem~\ref{thm:adhm-darboux-density-equivalence}.\footnote{%
		The full Darboux connection $\widehat{\mathcal A}_{\mathrm{DT}}$ has rank $r+3k$.
		Its restriction to the invariant subbundle
		spanned by the first $r+2k$ coordinate vectors
		defines $\mathcal A_{\mathrm{DT}}$.
		Thus $\mathsf G\in\GL(r+2k,\mathbb C)$,
		and the displayed block sizes are $k,r,k$.}

\end{enumerate}

This paper is organized as follows. Section~\ref{sec:darboux-transformations-and-densities} establishes the determinant formula for the Yang--Mills
density associated with a one-fold Darboux transformation.
Section~\ref{sec:line-solitons-and-grassmannian-parametrization}
constructs soliton solutions from arbitrary
Grassmannian points and shows how to construct
a smaller-size Yang matrix with the same
Yang--Mills density.
Section~\ref{sec:adhm} relates a one-fold Darboux
transformation to the ADHM construction and
recovers Osborn's formula as well as the associated gauge connection.
Section~\ref{sec:conclusion} summarizes the main results
and discusses possible extensions.
Finally, Appendix~\ref{app:onefold-nfold-darboux-correspondence}
establishes the correspondence between $n$-fold
DT for $\GL(r,\mathbb C)$ ASDYM equations
and one-fold DT for $\GL(nr,\mathbb C)$ ASDYM equations
at the level of Yang--Mills densities.
Appendix~\ref{app:proof} provides a detailed proof for Theorem~\ref{thm:adhm-darboux-connection-recovery}.

\section{Darboux transformations and Yang--Mills densities}
\label{sec:darboux-transformations-and-densities}

We begin our construction with a determinant formula for the Yang--Mills density associated with a one-fold Darboux transformation (DT) for the ASDYM equations. 

\begin{theorem}
	\label{thm:onefold-darboux-density}
	Let $\Lambda\in\GL(r,\mathbb C)$ be a constant spectral matrix, and let $\Theta$ be an invertible $r\times r$ matrix-valued function satisfying the vacuum linear system
	\begin{equation}
		\partial_z\Theta
		=(\partial_{\widetilde w}\Theta)\Lambda,
		\qquad
		\partial_w\Theta
		=(\partial_{\widetilde z}\Theta)\Lambda.
	\label{eq:vacuum-linear-system}
	\end{equation}
	Then $S=\Theta\Lambda\Theta^{-1}$ satisfies
	\begin{equation}
		\partial_z S
		=(\partial_{\widetilde w}S)S,
		\qquad
		\partial_w S
		=(\partial_{\widetilde z}S)S,
		\label{eq:onefold-darboux-miura}
	\end{equation}
	and hence solves the Yang equation \eqref{eq:yang-equation}.
	The corresponding Yang--Mills density is given by
	\begin{equation}
		\mathcal L[S]
		=-\bigl(
		\partial_z\partial_{\widetilde z}
		-\partial_w\partial_{\widetilde w}
		\bigr)^2\ln \det\Theta.
		\label{eq:onefold-darboux-density}
	\end{equation}
\end{theorem}

\begin{proof}
	Differentiating $S$ and using \eqref{eq:vacuum-linear-system},
	we obtain
	\[
	(\partial_z S)S^{-1}
	=(\partial_z\Theta)\Theta^{-1}-S(\partial_z\Theta)\Lambda^{-1}\Theta^{-1}
	=(\partial_{\widetilde w}\Theta)\Lambda\Theta^{-1}-S(\partial_{\widetilde w}\Theta)\Theta^{-1}
	=\partial_{\widetilde w}S.
	\]
	Similarly, 
	$
	(\partial_wS)S^{-1}
	=\partial_{\widetilde z}S.
	$
	Consequently, we have
	\[
	\partial_{\widetilde z}\bigl((\partial_zS)S^{-1}\bigr)-\partial_{\widetilde w}\bigl((\partial_wS)S^{-1}\bigr)
	=\partial_{\widetilde z}\partial_{\widetilde w}S-\partial_{\widetilde w}\partial_{\widetilde z}S
	=0,
	\]
	which proves that $S$ satisfies
	the Yang equation \eqref{eq:yang-equation}.
	
	As for the Yang--Mills density formula,
	using \eqref{eq:ym-density-yang-matrix} and the relations
	$(\partial_zS)S^{-1}
	=\partial_{\widetilde w}S$
	and
	$(\partial_wS)S^{-1}
	=\partial_{\widetilde z}S$, we obtain
	\begin{equation}
	\mathcal L[S]
	=
	\tr\Bigl[
	\bigl(
	\partial_{\widetilde z}\partial_{\widetilde w}S
	\bigr)^2
	-
	(\partial_{\widetilde w}^2S)
	(\partial_{\widetilde z}^2S)
	\Bigr].
	\label{eq:onefold-density-trace}
	\end{equation}
	Set
	$
	U:=(\partial_{\widetilde w}\Theta)\Theta^{-1}.
	$
	Direct differentiation using \eqref{eq:vacuum-linear-system} gives
	\[
	\partial_zU=(\partial_{\widetilde w}U)S,
	\qquad
	\partial_wU=(\partial_{\widetilde z}U)S,
	\qquad
	\partial_{\widetilde w}S=US-SU.
	\]
	Moreover, \eqref{eq:vacuum-linear-system} implies
	$
	\partial_z\partial_{\widetilde z}\Theta
	=
	\partial_w\partial_{\widetilde w}\Theta.
	$
	Using Jacobi's formula $\partial\ln\det\Theta=\tr[(\partial\Theta)\Theta^{-1}]$ and the cyclicity of the trace yield
	\[
	\begin{aligned}
		\bigl(
		\partial_z\partial_{\widetilde z}
		-\partial_w\partial_{\widetilde w}
		\bigr)\ln\det\Theta
		&=
		\tr\Bigl[
		U(\partial_w\Theta)\Theta^{-1}
		-
		(\partial_{\widetilde z}\Theta)\Theta^{-1}
		(\partial_z\Theta)\Theta^{-1}
		\Bigr]
		\\
		&=
		\tr\Bigl[
		U\Bigl(
		(\partial_{\widetilde z}\Theta)\Lambda\Theta^{-1}
		-
		S(\partial_{\widetilde z}\Theta)\Theta^{-1}
		\Bigr)
		\Bigr]
		=
		\tr\bigl(U\partial_{\widetilde z}S\bigr).
	\end{aligned}
	\]
	Using the identities for $U$ and $S$, we calculate
	\begin{equation}
	\begin{aligned}
		\partial_z\tr\bigl(U\partial_{\widetilde z}S\bigr)
		&=
		\tr\Bigl[
		(\partial_{\widetilde w}U)S
		(\partial_{\widetilde z}S)
		+
		U(\partial_{\widetilde z}\partial_{\widetilde w}S)S
		+
		U(\partial_{\widetilde w}S)
		(\partial_{\widetilde z}S)
		\Bigr]
		\\
		&=
		\partial_{\widetilde w}
		\tr\bigl(US\partial_{\widetilde z}S\bigr)
		+
		\tr\Bigl[
		(SU-US)
		(\partial_{\widetilde z}\partial_{\widetilde w}S)
		\Bigr]
		\\
		&=
		\partial_{\widetilde w}
		\tr\bigl(US\partial_{\widetilde z}S\bigr)
		-
		\tr\Bigl[
		(\partial_{\widetilde w}S)
		(\partial_{\widetilde z}\partial_{\widetilde w}S)
		\Bigr].
	\end{aligned}
	\label{eq:step1}
	\end{equation}
	Similarly, one obtains
	\begin{equation}
	\begin{aligned}
		\partial_w\tr\bigl(U\partial_{\widetilde z}S\bigr)
		&=
		\tr\Bigl[
		(\partial_{\widetilde z}U)S
		(\partial_{\widetilde z}S)
		+
		U(\partial_{\widetilde z}^2S)S
		+
		U(\partial_{\widetilde z}S)^2
		\Bigr]
		\\
		&=
		\partial_{\widetilde z}
		\tr\bigl(US\partial_{\widetilde z}S\bigr)
		-
		\tr\Bigl[
		(\partial_{\widetilde w}S)
		(\partial_{\widetilde z}^2S)
		\Bigr].
	\end{aligned}
	\label{eq:step2}
	\end{equation}
	
	Applying $\partial_{\widetilde z}$ to \eqref{eq:step1}
	and $\partial_{\widetilde w}$ to \eqref{eq:step2}, and subtracting,
	we find that the mixed derivative terms
	$\tr(US\partial_{\widetilde z}S)$ cancel. Thus
	\[
	\begin{aligned}
		\bigl(
		\partial_z\partial_{\widetilde z}
		-\partial_w\partial_{\widetilde w}
		\bigr)^2\ln\det\Theta
		&=
		\partial_{\widetilde w}
		\tr\Bigl[
		(\partial_{\widetilde w}S)
		(\partial_{\widetilde z}^2S)
		\Bigr]
		-
		\partial_{\widetilde z}
		\tr\Bigl[
		(\partial_{\widetilde w}S)
		(\partial_{\widetilde z}\partial_{\widetilde w}S)
		\Bigr]
		\\
		&=
		\tr\Bigl[
		(\partial_{\widetilde w}^2S)
		(\partial_{\widetilde z}^2S)
		-
		\bigl(
		\partial_{\widetilde z}\partial_{\widetilde w}S
		\bigr)^2
		\Bigr]
		=-\mathcal L[S],
	\end{aligned}
	\]
	where the second equality follows from the product rule
	and the commutativity of mixed partial derivatives.
	This proves \eqref{eq:onefold-darboux-density}.
\end{proof}

Theorem~\ref{thm:onefold-darboux-density} gives the Yang matrix
obtained by a one-fold DT and expresses its
Yang--Mills density in terms of a determinant.
We therefore define the associated ASDYM $\tau$-function by $\det\Theta$.
In Appendix~\ref{app:onefold-nfold-darboux-correspondence},
we extend this result to the $n$-fold DTs for
$\GL(r,\mathbb C)$ ASDYM equations
\cite{GilsonHamanakaHuangNimmo2020}
and show that each such construction can be
represented by a one-fold DT for
$\GL(nr,\mathbb C)$ ASDYM equations with the same Yang--Mills density.
Therefore, we focus on the construction 
in Theorem~\ref{thm:onefold-darboux-density}, which serves as
a common starting point for our subsequent study
of the Grassmannian structure of ASDYM
solitons and the ADHM construction of instantons.

\section{Solitons and Grassmannian parametrization}
\label{sec:line-solitons-and-grassmannian-parametrization}

In Theorem~\ref{thm:onefold-darboux-density}, we expressed the Yang--Mills
density in terms of the ASDYM $\tau$-function.
To construct ASDYM solitons, we now consider matrices
$\Theta$ built from finite linear combinations of
exponential functions, as in the investigation
of KP solitons.

\subsection{Grassmannian parametrization of ASDYM
	\texorpdfstring{$\tau$}{tau}-functions}
\label{subsec:grassmannian-tau-functions}

Recall that KP $\tau$-functions are parametrized
by ordered real spectral parameters
$\kappa_1<\cdots<\kappa_M$ and a point
$[A]\in\Gr(N,M;\mathbb R)$.
We establish an analogous Grassmannian parametrization
for ASDYM $\tau$-function by showing that, for any distinct nonzero
spectral parameters $\kappa_1,\ldots,\kappa_M\in\mathbb C$
and any $[A]\in\Gr(N,M;\mathbb C)$, one can always construct
a matrix $\Theta$ whose determinant agrees,
up to a nonzero constant factor, with the
$\tau$-function associated with $A$ in
\eqref{eq:kp-wronskian-tau}.

\begin{theorem}
	\label{thm:grassmannian-parametrization}
	Fix integers $1\leq N<M$, pairwise distinct nonzero parameters
	$\kappa_1,\ldots,\kappa_M\in\mathbb C$, and set
	$\Lambda=\diag(\kappa_1,\ldots,\kappa_M)$.
	For nonzero constants $\alpha_j,\beta_j\in\mathbb C$, define
	\begin{equation}
		\theta_j
		:=
		\alpha_j\kappa_j z+\kappa_j\beta_j w
		+\beta_j\widetilde z+\alpha_j\widetilde w,
		\qquad
		E:=\diag(e^{\theta_1},\ldots,e^{\theta_M}).
		\label{eq:asdym-exponential-phases}
	\end{equation}
	Let $A\in\mathbb C^{N\times M}$ have rank $N$.
	Define
	\begin{equation}
		\Theta
		:=
		\begin{pmatrix}
			AE\\
			B
		\end{pmatrix},
		\qquad
		B:=
		\left(
		\frac{\kappa_j^{M-N-i}}
		{\prod_{k\neq j}(\kappa_j-\kappa_k)}
		\right)_{1\leq i\leq M-N,1\leq j\leq M}.
		\label{eq:grassmannian-darboux-matrix}
	\end{equation}
	Then $\Theta$ satisfies \eqref{eq:vacuum-linear-system} and
	\begin{equation}
		\prod_{1\leq i<j\leq M}
		(\kappa_j-\kappa_i)\det\Theta
		=
		\det(AEV^T)
		=:
		\tau,
		\label{eq:grassmannian-wronskian-tau}
	\end{equation}
	where $V=(\kappa_j^{i-1})_{1\leq i\leq N,\,
		1\leq j\leq M}$.
	On each domain where $\tau\neq0$, the matrix
	$S=\Theta\Lambda\Theta^{-1}$ solves the Yang equation
	\eqref{eq:yang-equation},
	and the corresponding Yang--Mills density is
	\begin{equation}
		\mathcal{L}
		=
		-\bigl(
		\partial_z\partial_{\widetilde z}
		-\partial_w\partial_{\widetilde w}
		\bigr)^2\ln\det\Theta
		=-\bigl(
		\partial_z\partial_{\widetilde z}
		-\partial_w\partial_{\widetilde w}
		\bigr)^2\ln\tau.
		\label{eq:grassmannian-tau-density}
	\end{equation}
	For fixed spectral and phase parameters, $\mathcal L$
	depends only on the Grassmannian point $[A]$.
\end{theorem}

\begin{proof}
	Since $B$ is constant and
	\[
	\partial_z e^{\theta_j}
	=\kappa_j\partial_{\widetilde w}e^{\theta_j},
	\qquad
	\partial_w e^{\theta_j}
	=\kappa_j\partial_{\widetilde z}e^{\theta_j},
	\]
	the matrix $\Theta$ satisfies \eqref{eq:vacuum-linear-system}.
	Introduce an $M\times M$ Vandermonde matrix
	\[
	K
	=
	\bigl(\kappa_j^{i-1}\bigr)_{1\leq i,j\leq M}
	=
	\begin{pmatrix}
		1 & 1 & \cdots & 1\\
		\kappa_1 & \kappa_2 & \cdots & \kappa_M\\
		\vdots & \vdots & & \vdots\\
		\kappa_1^{M-1} & \kappa_2^{M-1} & \cdots & \kappa_M^{M-1}
	\end{pmatrix},
	\]
	then the first $N$ rows of $K$ form $V$.
	Comparing coefficients of degree $M-1$ in the
	Lagrange interpolation formula gives
	\[
	\sum_{j=1}^{M}
	\frac{\kappa_j^{i-1}}
	{\prod_{k\neq j}(\kappa_j-\kappa_k)}
	=
	\begin{cases}
		0, & 1\leq i<M,\\
		1, & i=M.
	\end{cases}
	\]
	By the definition of $B$, this identity implies that $BV^T=0$ and
	\[
	\Theta K^T
	=
	\begin{pmatrix}
		AEK^T\\
		BK^T
	\end{pmatrix}
	=
	\left(
	\begin{array}{c|c}
		AEV^T & *\\
		\hline
		0 &
		\begin{matrix}
			1 & * & \cdots & *\\
			0 & 1 & \ddots & \vdots\\
			\vdots & \ddots & \ddots & *\\
			0 & \cdots & 0 & 1
		\end{matrix}
	\end{array}
	\right),
	\]
	where the lower-right block has size
	$(M-N)\times(M-N)$ and determinant $1$.
	Taking determinants therefore yields
	\[
	\det(AEV^T)
	=
	\det\Theta\,\det K^T
	=
	\prod_{1\leq i<j\leq M}
	(\kappa_j-\kappa_i)\det\Theta.
	\]
	
	Since $\det K^T$ is a nonzero constant,
	$\Theta$ is invertible wherever $\tau\neq0$.
	The Yang equation and the density formula then follow
	from Theorem~\ref{thm:onefold-darboux-density}.
	Finally, replacing $A$ by another representative of
	the same Grassmannian point multiplies $\tau$
	by a nonzero constant and hence leaves
	$\mathcal L$ unchanged.
\end{proof}

	Theorem~\ref{thm:grassmannian-parametrization}
	establishes an ASDYM--KP correspondence at the level of
	$\tau$-functions parametrized by Grassmannian points.
	For any $[A]\in\Gr(N,M;\mathbb R)$,
	we restrict this construction to the
	 real slice
	$(z,\widetilde z,w,\widetilde w)\in\mathbb R^4$
	and impose the following specialization
	of the phase parameters and coordinates:
	\begin{equation}
		\alpha_j=\kappa_j,
		\qquad
		\beta_j=\kappa_j^2,
		\qquad
		\kappa_j\in\mathbb{R},
		\qquad
		x=\widetilde w,
		\qquad
		y=z+\widetilde z,
		\qquad
		t=w.
		\label{eq:asdym-kp-coordinates}
	\end{equation}
	Under this specialization, the ASDYM phases
	\eqref{eq:asdym-exponential-phases}
	coincide with the KP phases
	\eqref{eq:kp-seeds-and-phases},
	and the associated ASDYM $\tau$-function
	reduces to the Wronskian form
	\eqref{eq:kp-wronskian-tau}
	of the KP-II equation.
	In particular, if
	$\kappa_1<\cdots<\kappa_M$ and
	$[A]\in\Gr_{\geq 0}(N,M;\mathbb R)$, the Cauchy--Binet expansion \eqref{eq:kp-cauchy-binet} shows $\tau>0$ globally on $\mathbb{R}^3$.
	It therefore determines a regular
	KP-II line-soliton solution by
	\begin{equation*}
		u(x,y,t)=2\partial_x^2\ln\tau(x,y,t)
		=
		2\partial_x^2\ln\det(AEV^T).
	\end{equation*}
	Thus, the Grassmannian parametrization
	of ASDYM $\tau$-functions encompasses
	the regular line-soliton $\tau$-functions
	of KP-II as a subfamily.

\subsection{Grammian representation}
\label{subsec:grammian}

As in KP line-soliton theory
\cite{ChakravartyKodama2009}, we restrict
attention to irreducible $A$ in reduced
row-echelon form (RREF). Here, $A$ is called
irreducible if every column is nonzero and every
row contains at least one nonzero entry apart from
its pivot. Indeed, a zero column can be removed
without changing the $\tau$-function, while a row
containing only its pivot can be eliminated by
factoring out the corresponding exponential and
rescaling the remaining coefficients. Since such
reductions leave the Yang--Mills density unchanged,
we assume throughout that $A$ is irreducible.

We first show that simultaneously permuting the columns
of $A$ and the associated spectral parameters leaves the
$\tau$-function unchanged.

\begin{lemma}
	\label{lem:pivot-column-permutation}
	Let $A\in\mathbb C^{N\times M}$ be irreducible in RREF, with pivot set
	$I=\{i_1<\cdots<i_N\}$.
	For any $M\times M$ permutation matrix $P$,
	define
	\[
	A'=AP,\qquad
	\Lambda'=P^{-1}\Lambda P,\qquad
	E'=P^{-1}EP,\qquad
	V'=VP,\qquad
	B'=BP.
	\]
	Then the associated $\tau$-function is unchanged.
	In particular, $P$ can be chosen to move
	the pivot columns to the first $N$ positions,
	while preserving the relative order within
	the pivot and nonpivot columns.
	The resulting matrix takes the form
	\[
		A'=
		\begin{pmatrix}
			1 & 0 & \cdots & 0 & * & \cdots & *\\
			0 & 1 & \cdots & 0 & * & \cdots & *\\
			\vdots & \vdots & \ddots & \vdots
			& \vdots & & \vdots\\
			0 & 0 & \cdots & 1 & * & \cdots & *
		\end{pmatrix}
		=(I_N\mid C),
		\qquad
		C\in\mathbb C^{N\times(M-N)}.
	\]
	Since $A$ is irreducible, every row and column
	of $C$ contains at least one nonzero entry.
\end{lemma}

\begin{proof}
	Since $P^{-1}=P^T$, we have
	\[
	\tau '
	=\det\bigl(A'E'(V')^T\bigr)
	=\det\bigl(AP(P^{-1}EP)P^TV^T\bigr)
	=\det(AEV^T)
	=\tau.
	\]
	The equality of the corresponding
	Yang--Mills densities follows immediately
	from the density formula \eqref{eq:grassmannian-tau-density}.
\end{proof}

By Lemma~\ref{lem:pivot-column-permutation},
the spectral parameters and their associated columns
may be relabeled without changing the $\tau$-function.
We may therefore assume, without loss of generality,
that the pivot columns occupy the first $N$ positions
and choose the representative
$A=(I_N\mid C)$.
Under this normalization, the $\tau$-function
admits a Cauchy-type Grammian representation
\cite{ChakravartyLewkowMaruno2010}.

We introduce the block notation
\begin{equation}
	\Lambda=\diag(\Lambda_1,\Lambda_2),
	\qquad
	E=\diag(E_1,E_2),
	\qquad
	B=(B_1\mid B_2),
	\qquad
	V=(V_1\mid V_2),
	\label{eq:grassmannian-block-decomposition}
\end{equation}
where $\Lambda_1$, $E_1$, and $V_1$ are
$N\times N$ matrices, while $\Lambda_2$, $E_2$,
and $B_2$ are $(M-N)\times(M-N)$ matrices.
Define the diagonal matrices
\[
	\begin{aligned}
		D_1:=\diag\left(
		\prod_{\substack{j=1\\j\neq i}}^N
		(\kappa_i-\kappa_j)
		\right)_{i=1}^N,\qquad
		D_2:=\diag\left(
		\prod_{j=1}^N
		(\kappa_{N+k}-\kappa_j)
		\right)_{k=1}^{M-N},
	\end{aligned}
\]
and the Cauchy matrix
\[
	G:=\left(
	\frac{1}{\kappa_{N+k}-\kappa_i}
	\right)_{1\leq k\leq M-N, \, 1\leq i\leq N}.
\]
Equivalently, $G$ is the unique solution
of the Sylvester equation
\[
	\Lambda_2G-G\Lambda_1
	=\mathbf1_{M-N}\mathbf1_N^T,
\]
where $\mathbf1_n$ denotes the
$n$-dimensional column vector of ones.
Lagrange interpolation gives
\[
	V_2^T(V_1^T)^{-1}=D_2GD_1^{-1}.
\]
Recalling that $\tau=\det(AEV^T)$,
a direct substitution yields
\[
\begin{aligned}
	AEV^T
	&=E_1V_1^T+CE_2V_2^T\\
	&=\left(
	I_N+CE_2V_2^T(V_1^T)^{-1}E_1^{-1}
	\right)E_1V_1^T\\
	&=\left(
	I_N+CE_2D_2GD_1^{-1}E_1^{-1}
	\right)E_1V_1^T\\
	&=D_1\left(
	I_N+D_1^{-1}CD_2E_2GE_1^{-1}
	\right)D_1^{-1}E_1V_1^T.
\end{aligned}
\]
Taking determinants gives  the Grammian representation
\begin{align}
	\tau=\det(E_1V_1^T)\,\widehat{\tau},
	\qquad
	\widehat{\tau}:=	\det(I_N+D_1^{-1}CD_2E_2GE_1^{-1}).
	\label{eq:grassmannian-grammian-tau}
\end{align}
Since
$(\partial_z\partial_{\widetilde z}
-\partial_w\partial_{\widetilde w})^2
\ln\det(E_1V_1^T)=0$,
the prefactor in this $\tau$-function does not contribute to
the Yang--Mills density.
Thus \eqref{eq:grassmannian-tau-density}
is determined entirely by this Grammian $\tau$-function, i.e.
\[
\mathcal{L}=
-\left(\partial_z\partial_{\widetilde z}
-\partial_w\partial_{\widetilde w}
\right)^2
\ln\det\Theta
=
-\left(\partial_z\partial_{\widetilde z}
-\partial_w\partial_{\widetilde w}
\right)^2
\ln\tau
=
-\left(\partial_z\partial_{\widetilde z}
-\partial_w\partial_{\widetilde w}
\right)^2
\ln\widehat\tau.
\]
This establishes a direct correspondence between the Wronskian $\tau$-functions considered here and the Grammian  $\tau$-functions studied in \cite{McLoughlin2026,LiMarunoZhang2026b}.

\subsection{Recovery of a smaller Yang's matrix}
\label{subsec:grassmannian-field-recovery}

Theorem~\ref{thm:grassmannian-parametrization}
associates a $\GL(M,\mathbb{C})$-valued Yang's matrix
$S=\Theta\Lambda\Theta^{-1}$ with each point
$[A]\in\Gr(N,M;\mathbb C)$ through
\eqref{eq:grassmannian-darboux-matrix}.
Although this construction always provides
the Yang matrix of size $M$, such a large matrix
is not necessarily required to reproduce the
same Yang--Mills density.
For irreducible $A$, we now construct an
$\SL(r,\mathbb C)$-valued Yang's matrix with the same
Yang--Mills density, where
$r\leq1+\min\{N,M-N\}$.

Throughout this subsection, we
retain the notation and assumptions of
Section~\ref{subsec:grammian}. Set
\begin{equation}
	H:=C\Lambda_2-\Lambda_1C,\qquad
	\delta:=\rank H\leq\min\{N,M-N\},\qquad
	r:=\delta+1.
	\label{eq:recovery-rank}
\end{equation}
The quantity $\delta$ depends only on $\Lambda$
and the Grassmannian point $[A]$.
Indeed, let $W=\operatorname{row}(A)$ and write
$W\Lambda=\{w\Lambda:w\in W\}$.
Since $A=(I_N\mid C)$, we have
$A\Lambda-\Lambda_1A=(0\mid H)$.
Block row operations therefore give
\[
\rank
\begin{pmatrix}
	A\\
	A\Lambda
\end{pmatrix}
=
\rank
\begin{pmatrix}
	I_N&C\\
	0&H
\end{pmatrix}
=N+\delta.
\]
Thus $\delta=\dim(W+W\Lambda)-\dim W$,
so the size $r=\delta+1$ is determined by
$\Lambda$ and $[A]$.

\begin{theorem}
	\label{thm:grassmannian-recovery}

	Choose a rank factorization $H=PQ$, where
	$P\in\mathbb C^{N\times\delta}$ and
	$Q\in\mathbb C^{\delta\times(M-N)}$
	are constant matrices, and define
	\[
		\Lambda_0:=
		\begin{pmatrix}
			\Lambda_1&0\\
			0&B_2\Lambda_2B_2^{-1}
		\end{pmatrix},
		\quad
		X:=
		\begin{pmatrix}
			P&0\\
			0&B_2D_2\mathbf1_{M-N}
		\end{pmatrix},
		\quad
		Y:=
		\begin{pmatrix}
			0&QE_2\\
			\mathbf1_N^TD_1^{-1}&0
		\end{pmatrix}.
	\]
	Then the $r\times r$ matrix-valued functions
	\begin{equation}
		\mathcal J:=I_r+Y\Theta^{-1}\Lambda_0^{-1}X,
		\qquad
		\mathcal K:=-Y\Theta^{-1}X
		\label{eq:recovery-JK-definition}
	\end{equation}
	satisfy $\det\mathcal J=1$ and 
	\begin{equation}
		-(\partial_z\mathcal J)\mathcal J^{-1}
		=\partial_{\widetilde w}\mathcal K,
		\qquad
		-(\partial_w\mathcal J)\mathcal J^{-1}
		=\partial_{\widetilde z}\mathcal K.
		\label{eq:recovery-JK-equations}
	\end{equation}
	Consequently, $\mathcal J$ solves the
	Yang equation and yields the same
	Yang--Mills density as $S$.
	Its matrix size satisfies
	\[
	r=\rank(C\Lambda_2-\Lambda_1C)+1.
	\]
\end{theorem}

\begin{proof}
	Under the block notation
	\eqref{eq:grassmannian-block-decomposition},
	the matrix $\Theta$ in \eqref{eq:grassmannian-darboux-matrix} takes the form
	\[
	\Theta=
	\begin{pmatrix}
		AE \\
		B
	\end{pmatrix}=
	\begin{pmatrix}
		E_1&CE_2\\
		B_1&B_2
	\end{pmatrix}.
	\]
	Since $BV^T=0$ and both $B_2$ and $V_1$
	are invertible, we have
	\[
	B_2^{-1}B_1=-V_2^T(V_1^T)^{-1}=-D_2GD_1^{-1}.
	\]
	Recall the Sylvester equation $\Lambda_2G-G\Lambda_1=\mathbf1_{M-N}\mathbf1_N^T$,
	it follows that
	\[
	\begin{aligned}
		B_1\Lambda_1-B_2\Lambda_2B_2^{-1}B_1
		&=B_2\bigl(
		(B_2^{-1}B_1)\Lambda_1
		-\Lambda_2(B_2^{-1}B_1)
		\bigr)\\
		&=B_2(\Lambda_2D_2GD_1^{-1}-D_2GD_1^{-1}\Lambda_1) \\
		&=B_2D_2\mathbf1_{M-N}
		\mathbf1_N^TD_1^{-1}.
	\end{aligned}
	\]
	Together with $H=PQ$, this gives
	\[
	\begin{aligned}
		\Theta\Lambda-\Lambda_0\Theta
		&=
		\begin{pmatrix}
			0& HE_2\\
			B_1\Lambda_1-B_2\Lambda_2B_2^{-1}B_1&0
		\end{pmatrix} \\
		&=
		\begin{pmatrix}
			0&PQE_2\\
			B_2D_2\mathbf1_{M-N}
			\mathbf1_N^TD_1^{-1}&0
		\end{pmatrix} 
		=XY.
	\end{aligned}
	\]
	Hence
	$S=\Theta\Lambda\Theta^{-1}
	=\Lambda_0+XY\Theta^{-1}$.
	
Recall that \eqref{eq:onefold-darboux-miura} in Theorem~\ref{thm:onefold-darboux-density}
implies 
\[
\partial_zS=(\partial_{\widetilde w}S)S,
\qquad
\partial_wS=(\partial_{\widetilde z}S)S.
\]
Since $\Lambda_0$ and $X$ are constant,
and $X$ has full column rank $r$,
substituting $S=\Lambda_0+XY\Theta^{-1}$
and cancelling $X$ gives
\[
	\begin{aligned}
		\partial_z(Y\Theta^{-1})
		&=\partial_{\widetilde w}(Y\Theta^{-1})
		(\Lambda_0+XY\Theta^{-1}),\\
		\partial_w(Y\Theta^{-1})
		&=\partial_{\widetilde z}(Y\Theta^{-1})
		(\Lambda_0+XY\Theta^{-1}).
	\end{aligned}
\]
	Multiplying on the right by $\Lambda_0^{-1}X$
	and using the definitions of
	$\mathcal J$ and $\mathcal K$ yields
	\[
	\begin{aligned}
		\partial_z\mathcal J
		&=\partial_{\widetilde w}(Y\Theta^{-1})
		X\mathcal J
		=-(\partial_{\widetilde w}\mathcal K)\mathcal J,\\
		\partial_w\mathcal J
		&=\partial_{\widetilde z}(Y\Theta^{-1})
		X\mathcal J
		=-(\partial_{\widetilde z}\mathcal K)\mathcal J.
	\end{aligned}
	\]
	Sylvester's determinant theorem gives
	\[
	\det\mathcal J
	=\det(I_M+\Lambda_0^{-1}XY\Theta^{-1})
	=\frac{\det S}{\det\Lambda_0}
	=\frac{\det\Lambda}{\det\Lambda_0}
	=1,
	\]
	since $\det\Lambda_0
	=\det\Lambda_1\det\Lambda_2=\det\Lambda$.
	Thus $\mathcal J$ is invertible, and the
	preceding derivative identities imply
	\eqref{eq:recovery-JK-equations}.
	Taking the $\widetilde z$-derivative of
	the first equation and the
	$\widetilde w$-derivative of the second
	shows that $\mathcal J$ satisfies the
	Yang equation.
	
	To compare the Yang--Mills densities,
	note that
	\begin{equation}
	\begin{aligned}
		(\partial_zS)S^{-1}
		&=X\partial_{\widetilde w}(Y\Theta^{-1}),
		&
		(\partial_wS)S^{-1}
		&=X\partial_{\widetilde z}(Y\Theta^{-1}),\\
		(\partial_z\mathcal J)\mathcal J^{-1}
		&=\partial_{\widetilde w}(Y\Theta^{-1})X,
		&
		(\partial_w\mathcal J)\mathcal J^{-1}
		&=\partial_{\widetilde z}(Y\Theta^{-1})X.
	\end{aligned}
	\label{eq:gauge-potentials}
	\end{equation}
	Since $X$ is constant, differentiation
	preserves these factorizations.
	Substituting into
	\eqref{eq:ym-density-yang-matrix}
	and applying
	$\tr_M(XUXV)=\tr_r(UXVX)$,
	for $U,V\in\mathbb C^{r\times M}$,
	to each term gives
	$\mathcal L[\mathcal J]=\mathcal L[S]$.
	Therefore,
	\[
	\mathcal L[\mathcal J]
	=\mathcal L[S]
	=-\bigl(
	\partial_z\partial_{\widetilde z}
	-\partial_w\partial_{\widetilde w}
	\bigr)^2\ln\tau,
	\]
	where the last equality follows from
	Theorem~\ref{thm:grassmannian-parametrization}.
\end{proof}

The correspondence of the Yang--Mills density associated with $S$ and $\mathcal{J}$ in Theorem~\ref{thm:grassmannian-recovery} can be extended to the level of gauge potentials.
Let $\mathcal A_S$ and $\mathcal A_{\mathcal J}$
denote the associated connections determined by $S$ and
$\mathcal J$, respectively.
Identity \eqref{eq:gauge-potentials} gives
\[
\begin{aligned}
	\mathcal A_S
	&=-X\bigl[
	\partial_{\widetilde w}(Y\Theta^{-1})\,\mathrm dz
	+\partial_{\widetilde z}(Y\Theta^{-1})\,\mathrm dw
	\bigr],\\
	\mathcal A_{\mathcal J}
	&=-\bigl[
	\partial_{\widetilde w}(Y\Theta^{-1})\,\mathrm dz
	+\partial_{\widetilde z}(Y\Theta^{-1})\,\mathrm dw
	\bigr]X.
\end{aligned}
\]
Since $X$ is constant and has full column rank $r$,
choose a constant matrix
$X_0\in\mathbb C^{M\times(M-r)}$ such that
$g=(X\mid X_0)\in\mathrm{GL}(M,\mathbb C)$.
Then $g^{-1}X=\binom{I_r}{0}$.
As $\mathrm dg=0$, a gauge transformation
gives the transformed connection and its curvature
in the form
\[
g^{-1}\mathrm dg+g^{-1}\mathcal A_Sg=
g^{-1}\mathcal A_Sg
=\begin{pmatrix}
	\mathcal A_{\mathcal J}&*\\
	0&0
\end{pmatrix},
\qquad
g^{-1}\mathcal F_Sg
=\begin{pmatrix}
	\mathcal F_{\mathcal J}&*\\
	0&0
\end{pmatrix},
\]
Since the off-diagonal curvature blocks do not
contribute to traces of curvature products, we obtain
\[
\mathcal L[S]=\mathcal L[\mathcal J],
\qquad
\tr_M(\mathcal F_S\wedge \mathcal F_S)
=\tr_r(\mathcal F_{\mathcal J}\wedge \mathcal F_{\mathcal J}).
\]
This gives a direct explanation of the equality
of the Yang--Mills densities and also identifies
the corresponding second Chern forms.

\section{Instantons and the ADHM construction}
\label{sec:adhm}

The ADHM construction provides an algebraic description of finite-action anti-self-dual $\SU(r)$ instantons on $\mathbb R^4$.
In this section, we construct a one-fold Darboux transformation from regular ADHM data and recover Osborn's formula for the Yang--Mills density.
We then construct an explicit gauge transformation of the resulting Darboux connection, showing that the ADHM connection arises as the induced connection on a rank-$r$ quotient bundle.
This also gives a direct explanation of the equality
of their Yang--Mills densities.

On the Euclidean real slice
$x=(x^1,x^2,x^3,x^4)\in\mathbb R^4$, set
\[
	z=\frac{x^1+ix^2}{\sqrt2},\qquad
	w=\frac{x^3+ix^4}{\sqrt2},\qquad
	\widetilde z=\bar z,\qquad
	\widetilde w=-\bar w.
\]
Introduce the complex coordinates
\begin{equation}
	u=z+\widetilde w,\qquad
	v=w+\widetilde z,\qquad
	\bar u=\widetilde z-w,\qquad
	\bar v=z-\widetilde w.
	\label{eq:adhm-complex-coordinates}
\end{equation}
The transformation from $x$ to $(u,v)$ is
orthogonal and orientation-preserving, with
$|u|^2+|v|^2=|x|^2$.
The Euclidean Laplacian then takes the form
\begin{equation}
	\Delta_x:=\sum_{\mu=1}^4\partial_{x^\mu}^2
	=2\bigl(\partial_z\partial_{\widetilde z}
	-\partial_w\partial_{\widetilde w}\bigr)
	=4\bigl(\partial_u\partial_{\bar u}
	+\partial_v\partial_{\bar v}\bigr).
	\label{eq:adhm-euclidean-laplacian}
\end{equation}

Fix integers $r\geq2$ and $k\geq1$,
where $r$ is the rank of the associated vector bundle
and $k$ is the instanton number. The ADHM data are
constant matrices
$\mathsf B_1,\mathsf B_2\in\mathbb C^{k\times k}$,
$\mathsf I\in\mathbb C^{k\times r}$, and
$\mathsf J\in\mathbb C^{r\times k}$ satisfying
\begin{equation}
	\begin{aligned}[c]
		[\mathsf B_1,\mathsf B_2]+\mathsf I\mathsf J&=0,\\
		[\mathsf B_1,\mathsf B_1^\dagger]
		+[\mathsf B_2,\mathsf B_2^\dagger]
		+\mathsf I\mathsf I^\dagger-\mathsf J^\dagger\mathsf J&=0.
	\end{aligned}
	\label{eq:adhm-equations}
\end{equation}
Set $m=r+2k$ and define
$\Delta=(\alpha\mid\beta)\in\mathbb C^{m\times2k}$, where
\begin{equation}
	\alpha=\begin{pmatrix}
		\mathsf B_1-uI_k\\
		\mathsf B_2-vI_k\\
		\mathsf J
	\end{pmatrix},\qquad
	\beta=\begin{pmatrix}
		-\mathsf B_2^\dagger+\bar vI_k\\
		\mathsf B_1^\dagger-\bar uI_k\\
		\mathsf I^\dagger
	\end{pmatrix}.
	\label{eq:adhm-matrix}
\end{equation}
The ADHM equations \eqref{eq:adhm-equations}
are equivalent to
\begin{equation}
	\alpha^\dagger\alpha=\beta^\dagger\beta=\mathsf R,
	\qquad \alpha^\dagger\beta=0,
	\label{eq:adhm-factorization}
\end{equation}
where
\begin{equation}
	\mathsf R=(\mathsf B_1^\dagger-\bar uI_k)(\mathsf B_1-uI_k)
	+(\mathsf B_2^\dagger-\bar vI_k)(\mathsf B_2-vI_k)
	+\mathsf J^\dagger\mathsf J.
	\label{eq:adhm-R}
\end{equation}
We assume that the ADHM data are regular,
namely $\mathsf R(x)>0$ for all $x\in\mathbb R^4$,
and write $f=\mathsf R^{-1}$.

Let $\mathsf U\in\mathbb C^{m\times r}$ be a
smooth orthonormal frame of $\ker\Delta^\dagger$,
satisfying
\begin{equation}
	\Delta^\dagger\mathsf U=0,\qquad
	\mathsf U^\dagger\mathsf U=I_r.
	\label{eq:adhm-normalized-kernel}
\end{equation}
The ADHM connection is
$\mathcal A_{\mathrm{ADHM}}=\mathsf U^\dagger\mathrm d\mathsf U$
\cite{AtiyahHitchinDrinfeldManin1978}.
It is anti-Hermitian, and the frame can be chosen
so that $\operatorname{tr}\mathcal A_{\mathrm{ADHM}}=0$.
Since $\Delta^\dagger\Delta
=\operatorname{diag}(\mathsf R,\mathsf R)$,
the orthogonal decomposition
$\mathbb C^m=\ker\Delta^\dagger\oplus\operatorname{im}\Delta$
yields the completeness relation
\begin{equation}
	\mathsf U\mathsf U^\dagger+\alpha f\alpha^\dagger
	+\beta f\beta^\dagger=I_m.
	\label{eq:adhm-completeness}
\end{equation}
Differentiating
$\mathsf U^\dagger\alpha=\mathsf U^\dagger\beta=0$
and using \eqref{eq:adhm-completeness},
we obtain the curvature
$\mathcal F_{\mathrm{ADHM}}
=\mathrm d\mathcal A_{\mathrm{ADHM}}
+\mathcal A_{\mathrm{ADHM}}\wedge\mathcal A_{\mathrm{ADHM}}$
in the standard form
\cite{DoreyHollowoodKhozeMattis2002,Hamanaka2003}:
\begin{equation*}
	\mathcal F_{\mathrm{ADHM}}
	=\mathrm d\mathsf U^\dagger
	(I_m-\mathsf U\mathsf U^\dagger)\wedge\mathrm d\mathsf U
	=\mathsf U^\dagger\bigl(
	\mathrm d\alpha\,f\wedge\mathrm d\alpha^\dagger
	+\mathrm d\beta\,f\wedge\mathrm d\beta^\dagger
	\bigr)\mathsf U.
\end{equation*}
With the normalization
$\mathcal L=-\frac14\tr(F_{\mu\nu}F^{\mu\nu})$,
Osborn's formula \cite{Osborn1981} gives
\begin{equation}
	\mathcal L_{\mathrm{ADHM}}
	=-\frac14\Delta_x^2\ln\det\mathsf R
	=\frac14\Delta_x^2\ln\det f.
	\label{eq:adhm-density}
\end{equation}

We first construct a one-fold Darboux transformation
whose Yang--Mills density formula reproduces the determinant expression
in \eqref{eq:adhm-density}.

\begin{theorem}
	\label{thm:adhm-darboux-density-equivalence}
	Let the ADHM data be regular and define
	\begin{equation}
		\Theta=\begin{pmatrix}
			I_m&\alpha\\
			-\alpha^\dagger&0_k
		\end{pmatrix},
		\qquad
		\Lambda=\begin{pmatrix}
			-I_m&0\\
			0&I_k
		\end{pmatrix}.
		\label{eq:adhm-darboux-data}
	\end{equation}
	Then $\Theta$ satisfies the vacuum equations
	\begin{equation}
		\partial_z\Theta=(\partial_{\widetilde w}\Theta)\Lambda,
		\qquad
		\partial_w\Theta=(\partial_{\widetilde z}\Theta)\Lambda,
		\label{eq:adhm-darboux-vacuum-system}
	\end{equation}
	and $\det\Theta=\det\mathsf R>0$.
	The one-fold Darboux transformation gives
	a globally smooth Yang's matrix
	\begin{equation}
		S=\Theta\Lambda\Theta^{-1}
		=\begin{pmatrix}
			2\alpha f\alpha^\dagger-I_m&2\alpha f\\
			0&-I_k
		\end{pmatrix},
		\qquad S^{-1}=S,
		\label{eq:adhm-darboux-yang-matrix}
	\end{equation}
	with Yang--Mills density
	\begin{equation}
		\mathcal L_{\mathrm{DT}}
		=-\bigl(\partial_z\partial_{\widetilde z}
		-\partial_w\partial_{\widetilde w}\bigr)^2\ln\det\Theta
		=-\frac14\Delta_x^2\ln\det\mathsf R
		=\mathcal L_{\mathrm{ADHM}}.
		\label{eq:adhm-darboux-density-equivalence}
	\end{equation}
\end{theorem}

\begin{proof}
	The coordinate relations \eqref{eq:adhm-complex-coordinates} imply
	\[
	\partial_z\alpha=\partial_{\widetilde w}\alpha,\qquad
	\partial_w\alpha=\partial_{\widetilde z}\alpha,\qquad
	\partial_z\alpha^\dagger=-\partial_{\widetilde w}\alpha^\dagger,\qquad
	\partial_w\alpha^\dagger=-\partial_{\widetilde z}\alpha^\dagger.
	\]
	Substitution into \eqref{eq:adhm-darboux-data} proves
	\eqref{eq:adhm-darboux-vacuum-system}.
	Since $\alpha^\dagger\alpha=\mathsf R$, the block row operation
	\[
		\begin{pmatrix}
		I_m&0\\
		\alpha^\dagger&I_k
	\end{pmatrix}
	\Theta
	=
	\begin{pmatrix}
		I_m&0\\
		\alpha^\dagger&I_k
	\end{pmatrix}
	\begin{pmatrix}
		I_m&\alpha\\
		-\alpha^\dagger&0_k
	\end{pmatrix}
	=
	\begin{pmatrix}
		I_m&\alpha\\
		0&\mathsf R
	\end{pmatrix}
	\]
	gives $\det\Theta=\det\mathsf R>0$.
	Moreover, direct multiplication gives
	\[
	\Theta^{-1}=\begin{pmatrix}
		I_m-\alpha f\alpha^\dagger&-\alpha f\\
		f\alpha^\dagger&f
	\end{pmatrix},
	\]
	which yields \eqref{eq:adhm-darboux-yang-matrix}.
	In particular, $S$ and its associated connection are smooth
	on all of $\mathbb R^4$.
	Theorem~\ref{thm:onefold-darboux-density}
	and \eqref{eq:adhm-euclidean-laplacian} then give \eqref{eq:adhm-darboux-density-equivalence}.
\end{proof}

We now relate the corresponding gauge connections
by an explicit calculation.
Write
\[
\mathrm d:=\partial+\widetilde\partial,
\qquad
\partial:=\mathrm dz\,\partial_z+\mathrm dw\,\partial_w,
\qquad
\widetilde\partial
:=
\mathrm d\widetilde z\,\partial_{\widetilde z}
+\mathrm d\widetilde w\,\partial_{\widetilde w}.
\]
The Yang matrix \eqref{eq:adhm-darboux-yang-matrix}
defines the Darboux connection
\[
	\widehat{\mathcal A}_{\mathrm{DT}}
	=-(\partial_zS)S^{-1}\,\mathrm dz
	-(\partial_wS)S^{-1}\,\mathrm dw
	=-(\partial S)S,
\]
where we have used $S^{-1}=S$.
Let
\[
\mathsf H:=2\alpha f\alpha^\dagger-I_m
\]
denote the upper-left block of $S$.
Since $\mathsf H^2=I_m$, direct block multiplication gives
\[
\widehat{\mathcal A}_{\mathrm{DT}}
=
\begin{pmatrix}
	-(\partial\mathsf H)\mathsf H&
	2\partial(\alpha f)-2(\partial\mathsf H)\alpha f\\
	0&0
\end{pmatrix}.
\]
The upper-left block therefore defines a rank-$m$ reduced 
connection
\[
\mathcal A_{\mathrm{DT}}
:=-(\partial\mathsf H)\mathsf H,
\qquad
\mathsf{H}\in\GL(m,\mathbb{C}).
\]
Moreover, the upper-left block of the Yang equation
for $S$ is exactly
\[
\partial_{\widetilde z}
\bigl((\partial_z\mathsf H)\mathsf H^{-1}\bigr)
-
\partial_{\widetilde w}
\bigl((\partial_w\mathsf H)\mathsf H^{-1}\bigr)
=0.
\]

\begin{theorem}
	\label{thm:adhm-darboux-connection-recovery}
	Under the assumptions of
	Theorem~\ref{thm:adhm-darboux-density-equivalence},
	define
	\begin{equation}
		\mathsf G
		:=
		\begin{pmatrix}
			\alpha-\beta&\mathsf U&\alpha f
		\end{pmatrix}.
		\label{eq:adhm-darboux-gauge}
	\end{equation}
	Then $\mathsf G\in\GL(m,\mathbb C)$ is smooth
	on $\mathbb R^4$, with inverse
	\begin{equation}
		\mathsf G^{-1}
		=
		\begin{pmatrix}
			-f\beta^\dagger\\
			\mathsf U^\dagger\\
			\alpha^\dagger+\beta^\dagger
		\end{pmatrix}.
		\label{eq:adhm-darboux-gauge-inverse}
	\end{equation}
	Its action on the reduced Darboux connection is
	\begin{equation}
		\mathsf G^{-1}
		{\mathcal A}_{\mathrm{DT}}\mathsf G
		+\mathsf G^{-1}\mathrm d\mathsf G
		=
		\begin{pmatrix}
			0&
			f\,\mathrm d\beta^\dagger\,\mathsf U&
			-f\beta^\dagger\,\mathrm d\beta\,f\\
			0&
			\mathcal A_{\mathrm{ADHM}}&
			\mathsf U^\dagger\,\mathrm d\beta\,f\\
			0&0&0
		\end{pmatrix},
		\label{eq:adhm-darboux-transformed-connection}
	\end{equation}
	where the block sizes are $k,r,k$ and
	$\mathcal A_{\mathrm{ADHM}}
	=\mathsf U^\dagger\mathrm d\mathsf U$.
\end{theorem}

\begin{proof}
	The result follows from a direct computation,
	as detailed in Appendix~\ref{app:proof}.
\end{proof}

The block form in
\eqref{eq:adhm-darboux-transformed-connection}
also implies that
$\mathcal L_{\mathrm{DT}}=\mathcal L_{\mathrm{ADHM}}$.
It provides an alternative proof of \eqref{eq:adhm-darboux-density-equivalence} without using \eqref{eq:onefold-darboux-density}.

\begin{example}
	\label{ex:adhm-bpst}
	We show how the BPST instanton density \cite{BelavinPolyakovSchwartzTyupkin1975} is recovered in this Darboux-transformation setting.
	Take $k=1$, $r=2$, and
	\[
	\mathsf B_1=\mathsf B_2=0,
	\qquad
	\mathsf I=(\rho,0),
	\qquad
	\mathsf J=\begin{pmatrix}0\\\rho\end{pmatrix},
	\qquad \rho>0.
	\]
	These data satisfy \eqref{eq:adhm-equations}, and
	\[
	\alpha=
	\begin{pmatrix}-u\\-v\\0\\\rho\end{pmatrix},
	\qquad
	\mathsf R=|u|^2+|v|^2+\rho^2=|x|^2+\rho^2.
	\]
	The corresponding one-fold DT is determined by
	\[
		\Theta=
		\begin{pmatrix}
			1&0&0&0&-u\\
			0&1&0&0&-v\\
			0&0&1&0&0\\
			0&0&0&1&\rho\\
			\bar u&\bar v&0&-\rho&0
		\end{pmatrix},
		\qquad
		\Lambda=\diag(-1,-1,-1,-1,1).
	\]
	Thus $\det\Theta=|x|^2+\rho^2$, and
	\begin{equation}
		\mathcal L_{\mathrm{BPST}}
		=-\frac14\Delta_x^2\ln(|x|^2+\rho^2)
		=\frac{24\rho^4}{(|x|^2+\rho^2)^4}.
		\label{eq:adhm-bpst-density}
	\end{equation}
\end{example}

\section{Concluding remarks}
\label{sec:conclusion}

In this paper, we have investigated soliton solutions
and ADHM instantons of the ASDYM equations
within a unified framework based on the one-fold
Darboux transformation and its associated
$\tau$-functions, with particular emphasis on
their Yang--Mills Lagrangian densities.
For soliton solutions, we constructed
$\tau$-functions parametrized by arbitrary
points of the complex Grassmannian,
in analogy with the Grassmannian description
of KP solitons.
For ADHM instantons, we constructed a one-fold
Darboux transformation from regular ADHM data
and recovered Osborn's formula from the associated
determinant $\tau$-function.
We also gave an explicit gauge transformation
under which the resulting Darboux
connection becomes block upper triangular and
induces the ADHM connection on a rank-$r$
quotient bundle.

The connection between the ASDYM equations and the
KP hierarchy has long attracted considerable interest
\cite{MasonWoodhouse1996,KakeiIkedaTakasaki2002,Takasaki2026}.
One classical approach realizes the KP hierarchy
as a reduction of the ASDYM hierarchy
by allowing the gauge potentials to take values
in an infinite-dimensional algebra of
matrix pseudodifferential operators
\cite{AblowitzChakravartyTakhtajan1993}.
More recently, operator-free reductions of the ASDYM
equations with gauge group $\GL(2,\mathbb C)$
to the KP and mKP hierarchies have been established,
together with the construction of Gram-type solutions
\cite{LiMarunoZhang2026a}.
Related work has also expressed the Yang--Mills
Lagrangian density of ASDYM solutions in terms
of a $\tau$-function, which takes the form
of a Grammian KP $\tau$-function
for suitable specialization
\cite{LiMarunoZhang2026b}.

The main contribution of our work in this direction
is to establish a direct correspondence between soliton-type
ASDYM and KP $\tau$-functions through
a common determinant construction parametrized
by arbitrary points of the complex Grassmannian
$\Gr(N,M;\mathbb C)$ (Theorem~\ref{thm:grassmannian-parametrization}).
Since KP line-soliton solutions admit a complete
parametrization by finite Grassmannian, this correspondence allows us to apply
the systematic framework of
KP soliton theory to the study of ASDYM solitons.

In KP soliton theory, both $u$ and $\tau$-functions are scalar-valued, they are related by 
 $u=2\partial_x^2\ln\tau$.
In contrast, for the ASDYM equations, the Yang matrix is matrix-valued, whereas its associated
$\tau$-function is scalar-valued.
Consequently, expressing the Yang--Mills 
density in terms of $\tau$
does not directly provide an associated Yang's matrix.
Our second contribution is the explicit construction
of an  $\SL(r,\mathbb C)$-valued Yang's matrix
for each irreducible Grassmannian point
$[A]\in\Gr(N,M;\mathbb C)$, where
$r\leq 1+\min\{N,M-N\}$ (Theorem~\ref{thm:grassmannian-recovery}).
The resulting Yang's matrix has precisely the
Yang--Mills density determined
by the associated $\tau$-function.

Finally, we established the relation between
one-fold DT and the ADHM construction of instantons.
Osborn's formula for the instanton density and ADHM connection are recovered within this framework.
However, the corresponding spectral matrix $\Lambda=\diag(-I_m,I_k)$ has repeated eigenvalues, whereas our Grassmannian
parametrization requires pairwise distinct spectral parameters.
Consequently, the Grassmannian theory
developed in this paper does not directly extend
to ADHM instantons.
Extending this method to repeated
spectral parameters remains an interesting
direction for future research.

\section*{Acknowledgments}

This work was supported by the Japan Society for the
Promotion of Science (JSPS) under Grant No.~P25326
and the National Natural Science Foundation of China
(NSFC) under Grant No.~12601462.

\section*{Statements and Declarations}

\paragraph{Competing interests.}
The author declares no financial or non-financial competing
interests relevant to the content of this article.

\paragraph{Data availability.}
Data sharing is not applicable to this article, as no datasets
were generated or analysed during the current study.

\appendix

\section{Correspondence between one-fold DT and \texorpdfstring{$n$}{n}-fold DT}
\label{app:onefold-nfold-darboux-correspondence}

In this appendix, we retain the notation for the $n$-fold Darboux transformation
used in \cite[Section~3.1]{Li2026}.
Fix integers $n\geq1$ and $r\geq2$.
Let $\Lambda_i\in\GL(r,\mathbb C)$ be constant
matrices and let $\Theta_i$ be $r\times r$
matrix-valued functions satisfying the vacuum
linear system
\begin{equation}
	\partial_z\Theta_i
	=(\partial_{\widetilde w}\Theta_i)\Lambda_i,
	\qquad
	\partial_w\Theta_i
	=(\partial_{\widetilde z}\Theta_i)\Lambda_i,
	\qquad 1\leq i\leq n.
	\label{eq:nfold-darboux-vacuum-system}
\end{equation}
For $k\geq0$, set
$
\Pi_k:=
(\Theta_1\Lambda_1^k,\ldots,
\Theta_n\Lambda_n^k)
$
and form the $nr\times nr$ block matrix
\begin{equation}
	\Sigma:=
	\begin{pmatrix}
		\Theta_1 & \cdots & \Theta_n\\
		\Theta_1\Lambda_1 & \cdots & \Theta_n\Lambda_n\\
		\vdots & & \vdots\\
		\Theta_1\Lambda_1^{n-1}
		& \cdots & \Theta_n\Lambda_n^{n-1}
	\end{pmatrix}
	=
	\begin{pmatrix}
		\Pi_0\\
		\Pi_1\\
		\vdots\\
		\Pi_{n-1}
	\end{pmatrix}.
	\label{eq:nfold-darboux-block-matrix}
\end{equation}
We work locally on the domain where
$\det\Sigma\neq0$.
The corresponding monic Darboux matrix polynomial
is given by
\[
	T(\lambda)
	=\lambda^n I_r
	-\Pi_n\Sigma^{-1}
	\begin{pmatrix}
		I_r\\
		\lambda I_r\\
		\vdots\\
		\lambda^{n-1}I_r
	\end{pmatrix}
	=\lambda^n I_r+
	\sum_{k=0}^{n-1}T_k\lambda^k,
\]
where $T_k$ are $r\times r$ matrix-valued
functions. In particular, each coefficient $T_k$ admits the
quasideterminant representation
\[
	T_k=
	\left|
	\begin{array}{cccc}
		\Theta_1 & \cdots & \Theta_n & \delta_{0k}I_r\\
		\Theta_1\Lambda_1 & \cdots & \Theta_n\Lambda_n
		& \delta_{1k}I_r\\
		\vdots & & \vdots & \vdots\\
		\Theta_1\Lambda_1^{n-1}
		& \cdots & \Theta_n\Lambda_n^{n-1}
		& \delta_{n-1,k}I_r\\
		\Theta_1\Lambda_1^n
		& \cdots & \Theta_n\Lambda_n^n
		& \boxed{0}
	\end{array}
	\right|,
	\qquad 0\leq k\leq n-1,
\]
where $\delta_{jk}$ denotes the Kronecker delta
and the quasideterminant is expanded about the
lower-right block.

By construction, these coefficients
satisfy the interpolation identity
\[
	\Pi_n
	=-\sum_{k=0}^{n-1}T_k\Pi_k=-T_0\Pi_0-T_1\Pi_1-\cdots-T_{n-1}\Pi_{n-1}.
\]
This identity can be written in block-matrix form as
\begin{equation}
	\widehat{\Sigma}=S\Sigma,
	\qquad
	\widehat{\Sigma}:=
	\begin{pmatrix}
		\Pi_1\\
		\Pi_2\\
		\vdots\\
		\Pi_n
	\end{pmatrix},
	\qquad
	S:=
	\begin{pmatrix}
		0&I_r&&0\\
		0&0&\ddots&\\
		\vdots&&\ddots&I_r\\
		-T_0&-T_1&\cdots&-T_{n-1}
	\end{pmatrix}.
	\label{eq:darboux-companion}
\end{equation}
For $n=1$, the companion matrix reduces to
$S=-T_0$.

Now set
$
\Gamma:=
\diag(\Lambda_1,\ldots,\Lambda_n).
$
Since $\Pi_{k+1}=\Pi_k\Gamma$, we also have
$\widehat{\Sigma}=\Sigma\Gamma$.
Consequently, $S=\Sigma\Gamma\Sigma^{-1}.$
Moreover, \eqref{eq:nfold-darboux-vacuum-system} implies
\[
	\partial_z\Sigma
	=(\partial_{\widetilde w}\Sigma)\Gamma,
	\qquad
	\partial_w\Sigma
	=(\partial_{\widetilde z}\Sigma)\Gamma.
\]
Thus $\Sigma$ satisfies the same linear system
as in \eqref{eq:vacuum-linear-system}, with $\Gamma$ in
place of $\Lambda$.
By Theorem~\ref{thm:onefold-darboux-density}, the
companion matrix $S$ satisfies
\eqref{eq:onefold-darboux-miura} and hence the Yang equation
for $\GL(nr,\mathbb C)$.

Since $\Sigma$ and $\Gamma$ are invertible,
so is $S$, which in turn implies that $T_0$
is invertible. Comparing the lower-left
$r\times r$ blocks in \eqref{eq:onefold-darboux-miura},
we obtain
\begin{equation}
	-(\partial_zT_0)T_0^{-1}
	=\partial_{\widetilde w}T_{n-1},
	\qquad
	-(\partial_wT_0)T_0^{-1}
	=\partial_{\widetilde z}T_{n-1}.
	\label{eq:darboux-coefficient-miura}
\end{equation}
These relations show that
$T_0$ satisfies the
Yang equation \eqref{eq:yang-equation}.
Substituting \eqref{eq:darboux-coefficient-miura}
into \eqref{eq:ym-density-yang-matrix},
we obtain
\[
	\mathcal L[T_0]
	=
	\tr\Bigl[
	(\partial_{\widetilde z}
	\partial_{\widetilde w}T_{n-1})^2
	-
	(\partial_{\widetilde w}^2T_{n-1})
	(\partial_{\widetilde z}^2T_{n-1})
	\Bigr].
\]
On the other hand,
\eqref{eq:onefold-density-trace} gives
\[
\mathcal L[S]
=
\tr\Bigl[
(\partial_{\widetilde z}
\partial_{\widetilde w}S)^2
-
(\partial_{\widetilde w}^2S)
(\partial_{\widetilde z}^2S)
\Bigr].
\]
By the companion form \eqref{eq:darboux-companion},
for any $\mu,\nu\in\{\widetilde z,\widetilde w\}$,
we have
\[
\partial_\mu\partial_\nu S
=
-\begin{pmatrix}
	0&\cdots&0\\
	\vdots&&\vdots\\
	\partial_\mu\partial_\nu T_0
	&\cdots&
	\partial_\mu\partial_\nu T_{n-1}
\end{pmatrix}.
\]
Consequently, for any
$\mu,\nu,\rho,\sigma\in\{\widetilde z,\widetilde w\}$,
\[
\begin{aligned}
	(\partial_\mu\partial_\nu S)
	(\partial_\rho\partial_\sigma S)
	=
	\begin{pmatrix}
		0&\cdots&0\\
		\vdots&&\vdots\\
		(\partial_\mu\partial_\nu T_{n-1})
		(\partial_\rho\partial_\sigma T_0)
		&\cdots&
		(\partial_\mu\partial_\nu T_{n-1})
		(\partial_\rho\partial_\sigma T_{n-1})
	\end{pmatrix}.
\end{aligned}
\]
Taking the trace gives
\[
\tr\bigl[
(\partial_\mu\partial_\nu S)
(\partial_\rho\partial_\sigma S)
\bigr]
=
\tr\bigl[
(\partial_\mu\partial_\nu T_{n-1})
(\partial_\rho\partial_\sigma T_{n-1})
\bigr].
\]
In particular,
\[
\begin{aligned}
	\mathcal L[S]
	=
	\tr\Bigl[
	(\partial_{\widetilde z}
	\partial_{\widetilde w}T_{n-1})^2
	-
	(\partial_{\widetilde w}^2T_{n-1})
	(\partial_{\widetilde z}^2T_{n-1})
	\Bigr]
	=\mathcal L[T_0].
\end{aligned}
\]
Applying Theorem~\ref{thm:onefold-darboux-density}
to $S=\Sigma\Gamma\Sigma^{-1}$,
we conclude that
\begin{equation}
	\mathcal L[T_0]=\mathcal L[S]
	=-
	\bigl(
	\partial_z\partial_{\widetilde z}
	-\partial_w\partial_{\widetilde w}
	\bigr)^2\ln\det\Sigma.
	\label{eq:nfold-onefold-density}
\end{equation}

Thus, the $n$-fold DT for $\GL(r,\mathbb C)$
and the corresponding one-fold construction
for $\GL(nr,\mathbb C)$ have the same
Yang--Mills density.
This implies that we may study the Yang--Mills
densities arising from multi-fold DTs through the one-fold
construction of Theorem~\ref{thm:onefold-darboux-density}.

\section{Proof of Theorem~\ref{thm:adhm-darboux-connection-recovery}}
\label{app:proof}

	By \eqref{eq:adhm-factorization}
	and \eqref{eq:adhm-normalized-kernel}, a direct calculation yields
	\[
	\begin{pmatrix}
		-f\beta^\dagger\\
		\mathsf U^\dagger\\
		\alpha^\dagger+\beta^\dagger
	\end{pmatrix}
	\begin{pmatrix}
		\alpha-\beta&\mathsf U&\alpha f
	\end{pmatrix}
	=
	\begin{pmatrix}
		f\mathsf R&0&0\\
		0&I_r&0\\
		0&0&\mathsf Rf
	\end{pmatrix}
	=I_m.
	\]
	This proves \eqref{eq:adhm-darboux-gauge-inverse}.
	Regularity implies that $f=\mathsf R^{-1}$ is smooth.
	Since $\alpha$, $\beta$, and the chosen frame $\mathsf U$
	are smooth, so are $\mathsf G$ and $\mathsf G^{-1}$
	on $\mathbb R^4$.
	
	To verify \eqref{eq:adhm-darboux-transformed-connection}, we first establish some useful identities.
	The factorization and orthogonality relations give
	\begin{equation}
		\begin{aligned}
			\mathsf H\alpha
			&=2\alpha f(\alpha^\dagger\alpha)-\alpha
			=2\alpha f\mathsf R-\alpha
			=\alpha,\\
			\mathsf H\beta
			&=2\alpha f(\alpha^\dagger\beta)-\beta
			=-\beta,\\
			\mathsf H\mathsf U
			&=2\alpha f(\alpha^\dagger\mathsf U)-\mathsf U
			=-\mathsf U.
		\end{aligned}
		\label{eq:adhm-H-action}
	\end{equation}
	From \eqref{eq:adhm-matrix}
	and \eqref{eq:adhm-complex-coordinates}, we have
	\[
	\alpha+\beta
	=
	\begin{pmatrix}
		\mathsf B_1-\mathsf B_2^\dagger-2\widetilde wI_k\\
		\mathsf B_2+\mathsf B_1^\dagger-2\widetilde zI_k\\
		\mathsf J+\mathsf I^\dagger
	\end{pmatrix},
	\qquad
	\alpha-\beta
	=
	\begin{pmatrix}
		\mathsf B_1+\mathsf B_2^\dagger-2zI_k\\
		\mathsf B_2-\mathsf B_1^\dagger-2wI_k\\
		\mathsf J-\mathsf I^\dagger
	\end{pmatrix}.
	\]
	Hence
	\begin{equation}
		\partial(\alpha+\beta)=0,
		\qquad
		\widetilde\partial(\alpha-\beta)=0.
		\label{eq:adhm-alpha-beta-partial-identities}
	\end{equation}
	Differentiation gives
	\[
	\partial\alpha
	=
	\begin{pmatrix}
		-\mathrm dz\,I_k\\
		-\mathrm dw\,I_k\\
		0
	\end{pmatrix},
	\qquad
	\partial\alpha^\dagger
	=
	\begin{pmatrix}
		\mathrm dw\,I_k&-\mathrm dz\,I_k&0
	\end{pmatrix}.
	\]
	The coordinate relations
	\eqref{eq:adhm-complex-coordinates} then give
	\begin{equation}
		\begin{aligned}
			\mathrm d\beta
			&=
			\begin{pmatrix}
				\mathrm d\bar v\,I_k\\
				-\mathrm d\bar u\,I_k\\
				0
			\end{pmatrix}
			=
			\begin{pmatrix}
				(-\mathrm du+2\mathrm dz)I_k\\
				(-\mathrm dv+2\mathrm dw)I_k\\
				0
			\end{pmatrix}
			=\mathrm d\alpha-2\partial\alpha,\\
			\mathrm d\beta^\dagger
			&=
			\begin{pmatrix}
				\mathrm dv\,I_k&-\mathrm du\,I_k&0
			\end{pmatrix}
			=
			\begin{pmatrix}
				(2\mathrm dw+\mathrm d\bar u)I_k&
				(-2\mathrm dz+\mathrm d\bar v)I_k&
				0
			\end{pmatrix}
			=2\partial\alpha^\dagger-\mathrm d\alpha^\dagger.
		\end{aligned}
		\label{eq:adhm-alpha-beta-differential-identities}
	\end{equation}
	We also have
	\begin{equation}
		\begin{aligned}
			(\mathrm d\alpha^\dagger)\alpha
			&=-\mathrm d\bar u\,(\mathsf B_1-uI_k)
			-\mathrm d\bar v\,(\mathsf B_2-vI_k)\\
			&=
			\begin{pmatrix}
				-\mathsf B_2+vI_k&
				\mathsf B_1-uI_k&
				\mathsf I
			\end{pmatrix}
			\begin{pmatrix}
				\mathrm d\bar v\,I_k\\
				-\mathrm d\bar u\,I_k\\
				0
			\end{pmatrix}
			=\beta^\dagger\mathrm d\beta.
		\end{aligned}
		\label{eq:adhm-differential-pairing}
	\end{equation}
	Differentiating $\mathsf Rf=I_k$ gives
	\[
	(\mathrm d\mathsf R)f+\mathsf R\,\mathrm df=0.
	\]
	Since $\mathsf R=\alpha^\dagger\alpha$, this yields
	\begin{equation}
		\mathrm df
		=-f(\mathrm d\mathsf R)f
		=-f\bigl(
		(\mathrm d\alpha^\dagger)\alpha
		+\alpha^\dagger\mathrm d\alpha
		\bigr)f.
		\label{eq:adhm-f-differential}
	\end{equation}
	
	Set
	$\nabla_{\mathrm{DT}}
	:=\mathrm d+{\mathcal A}_{\mathrm{DT}}$.
	We now compute the three block columns of
	$\nabla_{\mathrm{DT}}\mathsf G$.
	
	\begin{enumerate}
		\item \textit{First block column.}
		Equation \eqref{eq:adhm-H-action} gives
		\[
		\mathsf H(\alpha-\beta)=\alpha+\beta,
		\qquad
		\mathsf H(\alpha+\beta)=\alpha-\beta.
		\]
		By the product rule and
		\eqref{eq:adhm-alpha-beta-partial-identities},
		\begin{equation}
			\begin{aligned}
				\nabla_{\mathrm{DT}}(\alpha-\beta)
				&=
				\mathrm d(\alpha-\beta)
				-(\partial\mathsf H)\mathsf H(\alpha-\beta)\\
				&=
				\mathrm d(\alpha-\beta)
				-(\partial\mathsf H)(\alpha+\beta)\\
				&=
				\mathrm d(\alpha-\beta)
				-\partial\bigl(\mathsf H(\alpha+\beta)\bigr)
				+\mathsf H\partial(\alpha+\beta)\\
				&=
				\mathrm d(\alpha-\beta)
				-\partial(\alpha-\beta)=
				\widetilde\partial(\alpha-\beta)
				=0.
			\end{aligned}
			\label{eq:adhm-gauge-first-column}
		\end{equation}
		
		\item \textit{Second block column.}
		Differentiating
		$\alpha^\dagger\mathsf U=\beta^\dagger\mathsf U=0$
		gives
		\[
		\alpha^\dagger\mathrm d\mathsf U
		=-(\mathrm d\alpha^\dagger)\mathsf U,
		\qquad
		\beta^\dagger\mathrm d\mathsf U
		=-(\mathrm d\beta^\dagger)\mathsf U.
		\]
		The completeness relation
		\eqref{eq:adhm-completeness} therefore gives
		\[
		\begin{aligned}
			\mathrm d\mathsf U
			&=
			\bigl(
			\mathsf U\mathsf U^\dagger
			+\alpha f\alpha^\dagger
			+\beta f\beta^\dagger
			\bigr)\mathrm d\mathsf U\\
			&=
			\mathsf U(\mathsf U^\dagger\mathrm d\mathsf U)
			-\alpha f(\mathrm d\alpha^\dagger)\mathsf U
			-\beta f(\mathrm d\beta^\dagger)\mathsf U.
		\end{aligned}
		\]
		Since $\mathsf H\mathsf U=-\mathsf U$
		and $\alpha^\dagger\mathsf U=0$,
		\[
		\begin{aligned}
			{\mathcal A}_{\mathrm{DT}}\mathsf U
			&=
			-(\partial\mathsf H)\mathsf H\mathsf U
			=(\partial\mathsf H)\mathsf U\\
			&=
			2\partial(\alpha f\alpha^\dagger)\mathsf U\\
			&=
			2(\partial\alpha)f\alpha^\dagger\mathsf U
			+2\alpha(\partial f)\alpha^\dagger\mathsf U
			+2\alpha f(\partial\alpha^\dagger)\mathsf U\\
			&=
			2\alpha f(\partial\alpha^\dagger)\mathsf U.
		\end{aligned}
		\]
		Using
		\eqref{eq:adhm-alpha-beta-differential-identities}
		and
		$\mathcal A_{\mathrm{ADHM}}
		=\mathsf U^\dagger\mathrm d\mathsf U$,
		we obtain
		\begin{equation}
			\begin{aligned}
				\nabla_{\mathrm{DT}}\mathsf U
				&=
				\mathrm d\mathsf U
				+{\mathcal A}_{\mathrm{DT}}\mathsf U\\
				&=
				\mathsf U\mathcal A_{\mathrm{ADHM}}
				-\alpha f(\mathrm d\alpha^\dagger)\mathsf U
				-\beta f(\mathrm d\beta^\dagger)\mathsf U
				+2\alpha f(\partial\alpha^\dagger)\mathsf U\\
				&=
				\mathsf U\mathcal A_{\mathrm{ADHM}}
				+\alpha f
				(2\partial\alpha^\dagger-\mathrm d\alpha^\dagger)
				\mathsf U
				-\beta f(\mathrm d\beta^\dagger)\mathsf U\\
				&=
				(\alpha-\beta)f(\mathrm d\beta^\dagger)\mathsf U
				+\mathsf U\mathcal A_{\mathrm{ADHM}}.
			\end{aligned}
			\label{eq:adhm-gauge-middle-column}
		\end{equation}
		
		\item \textit{Third block column.}
		Applying $\partial$ to
		$\mathsf H\alpha=\alpha$
		in \eqref{eq:adhm-H-action} gives
		\[
		(\partial\mathsf H)\alpha
		+\mathsf H\partial\alpha
		=\partial\alpha,
		\]
		so
		\[
		(\partial\mathsf H)\alpha
		=(I_m-\mathsf H)\partial\alpha
		=2(I_m-\alpha f\alpha^\dagger)\partial\alpha.
		\]
		Thus
		\[
		\begin{aligned}
			\nabla_{\mathrm{DT}}(\alpha f)
			&=
			\mathrm d(\alpha f)
			-(\partial\mathsf H)\mathsf H\alpha f\\
			&=
			(\mathrm d\alpha)f+\alpha\,\mathrm df
			-(\partial\mathsf H)\alpha f\\
			&=
			(\mathrm d\alpha)f+\alpha\,\mathrm df
			-2(I_m-\alpha f\alpha^\dagger)(\partial\alpha)f\\
			&=
			(\mathrm d\alpha-2\partial\alpha)f
			+\alpha\,\mathrm df
			+2\alpha f\alpha^\dagger(\partial\alpha)f.
		\end{aligned}
		\]
		Substituting
		\eqref{eq:adhm-alpha-beta-differential-identities}
		and \eqref{eq:adhm-f-differential},
		and then applying \eqref{eq:adhm-differential-pairing},
		gives
		\[
		\begin{aligned}
			\nabla_{\mathrm{DT}}(\alpha f)
			&=
			(\mathrm d\beta)f
			-\alpha f
			\bigl(
			(\mathrm d\alpha^\dagger)\alpha
			+\alpha^\dagger\mathrm d\alpha
			\bigr)f
			+2\alpha f\alpha^\dagger(\partial\alpha)f\\
			&=
			(\mathrm d\beta)f
			-\alpha f
			\bigl(
			(\mathrm d\alpha^\dagger)\alpha
			+\alpha^\dagger
			(\mathrm d\alpha-2\partial\alpha)
			\bigr)f\\
			&=
			(\mathrm d\beta)f
			-\alpha f
			\bigl(
			\beta^\dagger\mathrm d\beta
			+\alpha^\dagger\mathrm d\beta
			\bigr)f\\
			&=
			(\mathrm d\beta)f
			-\alpha f(\alpha^\dagger+\beta^\dagger)
			(\mathrm d\beta)f.
		\end{aligned}
		\]
		By \eqref{eq:adhm-completeness},
		\[
		I_m-\alpha f\alpha^\dagger
		=\mathsf U\mathsf U^\dagger+\beta f\beta^\dagger.
		\]
		We therefore have
		\begin{equation}
			\begin{aligned}
				\nabla_{\mathrm{DT}}(\alpha f)
				&=
				(I_m-\alpha f\alpha^\dagger)(\mathrm d\beta)f
				-\alpha f\beta^\dagger(\mathrm d\beta)f\\
				&=
				\mathsf U\mathsf U^\dagger(\mathrm d\beta)f
				+\beta f\beta^\dagger(\mathrm d\beta)f
				-\alpha f\beta^\dagger(\mathrm d\beta)f\\
				&=
				-(\alpha-\beta)f\beta^\dagger(\mathrm d\beta)f
				+\mathsf U
				\bigl(\mathsf U^\dagger\mathrm d\beta\,f\bigr).
			\end{aligned}
			\label{eq:adhm-gauge-last-column}
		\end{equation}
	\end{enumerate}
	
	Equations
	\eqref{eq:adhm-gauge-first-column},
	\eqref{eq:adhm-gauge-middle-column}, and
	\eqref{eq:adhm-gauge-last-column} give
	\[
	\nabla_{\mathrm{DT}}\mathsf G
	=
	\mathsf G
	\begin{pmatrix}
		0&
		f(\mathrm d\beta^\dagger)\mathsf U&
		-f\beta^\dagger(\mathrm d\beta)f\\
		0&
		\mathcal A_{\mathrm{ADHM}}&
		\mathsf U^\dagger(\mathrm d\beta)f\\
		0&0&0
	\end{pmatrix}.
	\]
	Multiplying on the left by $\mathsf G^{-1}$ and using
	$\nabla_{\mathrm{DT}}\mathsf G
	=\mathrm d\mathsf G
	+{\mathcal A}_{\mathrm{DT}}\mathsf G$
	proves \eqref{eq:adhm-darboux-transformed-connection}.

\end{document}